\documentclass[10pt,twocolumn,english,aps,prl,showpacs,superscriptaddress,groupaddress,floatfix,graphics,graphicx]{revtex4-2}

\usepackage[T1]{fontenc}
\usepackage{color}
\usepackage[english]{babel}
\usepackage{float}
\usepackage{amsmath}
\usepackage{amssymb}
\usepackage{graphicx}
\usepackage{xcolor}
\usepackage{soul}
\usepackage[bookmarks]{hyperref}
\hypersetup{colorlinks,linkcolor=blue,citecolor=blue,urlcolor=blue}
\usepackage[all]{hypcap}

\usepackage{soul}
\usepackage{comment}
\usepackage{orcidlink}

\usepackage{amsmath}
\usepackage{bm}
\usepackage{siunitx}
\usepackage{nicefrac}

\usepackage{graphicx}
\usepackage{xcolor}
\usepackage{xspace}
\usepackage{float}

\newcommand{\ket}[1]{| {#1} \rangle} 
\newcommand{\bra}[1]{\langle {#1} |} 
\DeclareMathAlphabet\mathbfcal{OMS}{cmsy}{b}{n}

\usepackage{hyperref}
\newcommand{\SAVBA}{\affiliation{Institute of Informatics, Slovak Academy of Sciences, 84507 Bratislava, Slovakia}}
\newcommand{\FFBG}{\affiliation {Faculty of Physics, University of Belgrade, 11001 Belgrade, Serbia}}

\begin{document}
\title{Layer-selective chirality switch in bilayer graphene intercalated by Janus monolayers}

\author{Marko Milivojevi{\' c}} \SAVBA \FFBG
\date{\today}

\begin{abstract}
We predict that intercalating bilayer graphene with nonmagnetic 
WSSe or magnetic MnSSe Janus monolayers induces a layer-selective 
switch of the in-plane Rashba spin texture, resulting in opposite 
spin current directions in the top and bottom graphene layers. 
First-principles calculations reveal that both Janus monolayers 
decouple the two graphene layers while simultaneously inducing 
opposite signs of the proximity-induced Rashba spin-orbit 
coupling in each. Tight-binding modeling of the proximitized 
layers, combined with Rashba-Edelstein charge-to-spin conversion 
calculations, confirms that the spin current direction can be 
independently controlled by gating the top or bottom graphene 
layer. Bilayer graphene intercalated by Janus monolayers thus 
represents a promising platform for gate-tunable, layer-selective 
spintronic devices.
\end{abstract}
\maketitle

\section{Introduction}
Spintronics~\cite{ZFS04,FME+07} is a subfield of electronics which focuses on spin rather than charge as a carrier of information. Spintronics applications require materials with sizable spin-orbit coupling (SOC), magnetic ordering, or both. This requirement excludes many materials with high mobility and excellent transport properties, such as graphene~\cite{NGM+04}, from being useful in spintronics. However, progress in van der Waals (vdW) heterostructure fabrication~\cite{Geim2013} has enabled engineering of desired properties in target materials without destroying their intrinsic electronic structure. This is achieved by stacking the target material with 
functional layers whose properties are transferred via the proximity effect, enabling a wide range of vdW heterostructures 
as promising platforms for electronic and spintronic device applications~\cite{Liu2016, Novoselov2016, Liang2020, 
Sierra2021:NatNano, Zhao2025}.

A paradigmatic example is the spin-orbit proximity effect in graphene/transition-metal dichalcogenide (TMDC) heterostructures~\cite{Gmitra2015,Gmitra2016}, where the theoretically predicted proximity-induced SOC was experimentally confirmed through charge-to-spin conversion measurements~\cite{GCR17,GKB+19,HSI+20,KHA+20,HKZ+21,CSC+22}. The possibility to control and modify the spin texture in 
graphene using different substrates~\cite{MGK+24} and the 
twist angle~\cite{YMK+23} has enabled control of spin 
relaxation~\cite{Sierra2025:NatMater}, as well as the 
direction and magnitude of the spin 
current~\cite{Veneri2022,IGH+22,MMG26}. Beyond enabling spin current flow, practical spin logic devices also require control over the spin current direction. One recent proposal of this is the reversal of the spin current 
direction driven by switching the out-of-plane ferroelectric 
polarization of In$_2$Se$_3$~\cite{BKL+24} in the 
graphene/In$_2$Se$_3$ heterostructure~\cite{MMJ+26}.

Whereas the control of the spin current direction via ferroelectric switching relies on active manipulation of the ferroelectric material, the underlying physics requires control over the chirality of the Rashba-induced spin texture, whose sign determines the direction of the spin current flow. An alternative and more direct route is to engineer opposite spin textures in the two graphene layers simultaneously, and to activate them selectively using electrical gating. In this work, we reveal a layer-selective chirality switch in bilayer graphene intercalated by nonmagnetic WSSe (Gr/WSSe/Gr) and magnetic MnSSe (Gr/MnSSe/Gr) Janus monolayers, which serve as a spacer between the top and bottom graphene layers, efficiently decoupling them while simultaneously inducing opposite spin textures in each layer. As a consequence, the spin currents in the top and bottom graphene flow in opposite directions, enabling a current switch simply by gating the top or bottom graphene layer independently. We demonstrate this 
by performing charge-to-spin conversion calculations using an 
effective tight-binding model of the proximitized top and 
bottom graphene layers that incorporates the interaction with 
the intercalated spacer, providing a simple and efficient 
route for electrical control of the spin current direction 
and paving the way toward gate-tunable spintronic devices 
based on vdW heterostructures. Notably, the Janus 
WSSe monolayer has been experimentally synthesized~\cite{Lin2020, Harris2023}, making Gr/WSSe/Gr an 
experimentally accessible platform for the effects predicted 
in this work.

This paper is organized as follows. In Sec.~\ref{sec:STRUCT}, we 
present the structural details of the studied heterostructures 
and their DFT band structure. In Sec.~\ref{sec:TB}, we introduce 
the effective tight-binding model of proximitized graphene that 
captures the spin physics of the top and bottom graphene layers 
in both heterostructures. In Sec.~\ref{sec:charge-spin}, we 
outline the numerical procedure used to calculate the 
Rashba-Edelstein efficiency. Our main results are presented in 
Sec.~\ref{sec:results}, and our conclusions are summarized in Sec.~\ref{sec:conclusions}. 

\begin{figure}[t]
    \centering
\includegraphics[width=0.35\textwidth]{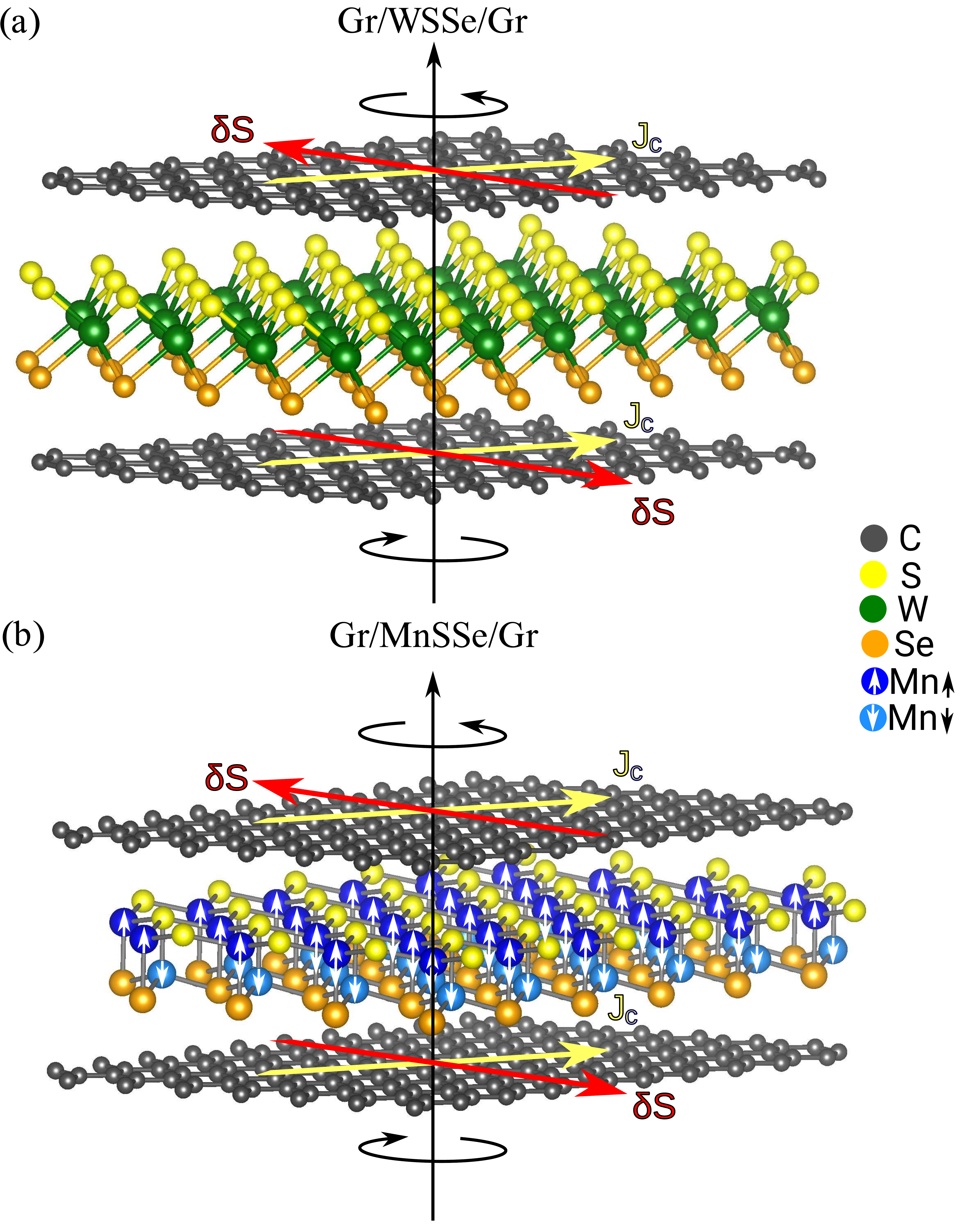}
    \caption{Schematic view of AA-stacked bilayer graphene intercalated by (a) WSSe (b) MnSSe Janus monolayer.
    Carbon (gray), sulfur (yellow), tungsten (dark green), and selenium (orange) atoms are shown. In panel (b), dark blue and light blue spheres represent Mn atoms with spin-up and spin-down magnetic moments, respectively, oriented perpendicular to the graphene plane in the model configuration considered here, reflecting the type-A antiferromagnetic ordering of MnSSe. The red circular arrows describe the chirality of the Rashba-induced spin texture in the top and bottom graphene layers, with clockwise/counterclockwise directions corresponding to positive/negative sign of the Rashba phase. As a direct consequence of the opposite chirality, a charge current ${\rm J}_{\rm c}$
    (blue arrows) generates spin accumulations $\delta S$ (red arrows) in opposite directions in the top and bottom graphene layers, enabling layer-selective control of the spin current direction.}
    \label{fig:structure}
\end{figure}
\section{Structural details and electronic structure}\label{sec:STRUCT}

We focus on two heterostructures: a nonmagnetic Gr/WSSe/Gr heterostructure, in which AA-stacked bilayer graphene is intercalated by a Janus WSSe monolayer (Fig.~\ref{fig:structure}(a)), and a magnetic Gr/MnSSe/Gr heterostructure, where the intercalating layer is the type-A Janus antiferromagnet MnSSe~\cite{SIC22} (Fig.~\ref{fig:structure}(b)). In their nonmagnetic phase, both WSSe and MnSSe have the ${\bf C}_{3{\rm v}}$ point group symmetry. The intercalation of a Janus monolayer breaks the horizontal mirror symmetry of AA-stacked bilayer graphene, while its three-fold rotational symmetry reduces the six-fold rotational symmetry of the bilayer to three-fold. The commensurate alignment between the bilayer graphene and the Janus monolayer preserves the common vertical mirror symmetry, resulting in ${\bf C}_{3{\rm v}}$ crystallographic symmetry of both heterostructures, which determines the symmetry-allowed 
spin physics studied here.

In the MnSSe case, the magnetic ordering must also be taken into account. MnSSe is a layered type-A antiferromagnet, in which the magnetic moments of the top and bottom Mn planes are aligned in 
opposite directions, as illustrated in Fig.~\ref{fig:structure}(b). 
Since the top and bottom graphene layers are in direct proximity to Mn planes with opposite magnetic orientations, the magnetic 
phase of MnSSe is expected to induce opposite proximity exchange interactions in the two graphene layers, breaking time-reversal symmetry~\cite{MG26}. Although the magnetocrystalline anisotropy of MnSSe 
theoretically favors an in-plane easy axis~\cite{SIC22}, 
out-of-plane antiferromagnetic ordering has been 
experimentally observed in the related MnSe monolayer~\cite{AHK+21}, contrary to theoretical 
predictions~\cite{SIC22}. This motivates us to consider 
the out-of-plane spin configuration as a model case to 
demonstrate the generality of the layer-selective chirality 
switch mechanism in the presence of both proximity-induced 
SOC and out-of-plane exchange interactions. To quantify the energy cost of this (out-of-plane) choice, we compare the total energies of the Gr/MnSSe/Gr heterostructure with in-plane and out-of-plane magnetic configurations,  finding an energy difference of only $0.19$~meV per Mn atom in 
favor of the in-plane configuration, consistent with~\cite{SIC22}. This small value confirms that the out-of-plane configuration 
is nearly degenerate with the ground state. Moreover, 
proximity effects from adjacent layers have been shown to 
reorient the magnetic anisotropy in two-dimensional magnets~\cite{Pozsar2026}, 
suggesting that placing the Gr/MnSSe/Gr heterostructure on a 
suitable substrate could provide the additional magnetic 
anisotropy needed to stabilize the out-of-plane spin 
configuration experimentally.

The electronic structure of the Gr/WSSe/Gr and Gr/MnSSe/Gr heterostructures was calculated using Density Functional Theory (DFT) as implemented in Q{\sc{uantum}} ESPRESSO~\cite{QE1,QE2} code, with van der Waals~\cite{G06,Barone2009} and dipole correction~\cite{B99} included. Spin-orbit coupling was treated using fully-relativistic 
pseudopotentials. Full computational details are given in 
Appendix~\ref{AppA}.
\begin{figure}[t]
    \centering
\includegraphics[width=0.48\textwidth]{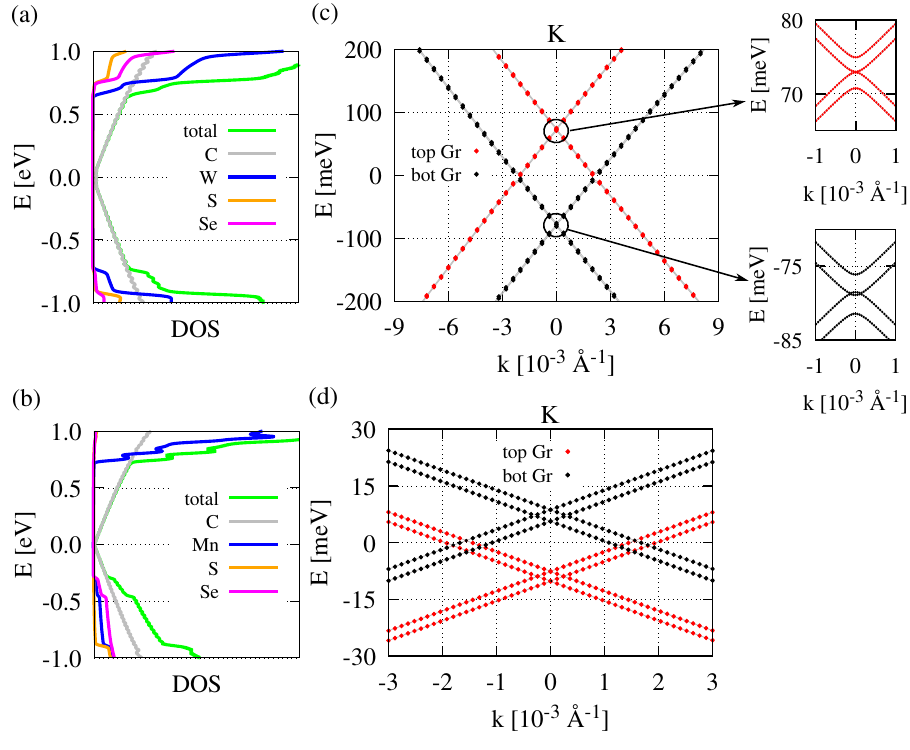}
    \caption{Atom-projected density of states of (a) Gr/WSSe/Gr and (b) Gr/MnSSe/Gr heterostructures. In both cases, the carbon states reside inside the semiconducting gap of the Janus monolayer, confirming the absence of hybridization between graphene and substrate states near the Fermi level. Layer-resolved fatband structure of (c) Gr/WSSe/Gr and (d) Gr/MnSSe/Gr in the vicinity of the K point, with the top (red) and bottom (black) graphene layer contributions given. The inset in (c) shows a zoom around the Dirac points of both the top and bottom graphene layers, showing spin splitting due to proximity-induced spin-orbit coupling.}    \label{fig:PROJWFC}
\end{figure}

Since both WSSe and MnSSe are semiconductors, whereas graphene is a semimetal, the electronic states around the Fermi level are expected to belong to graphene, albeit with modified properties due to the proximity of the Janus monolayers. To understand the low-energy electronic structure of the studied heterostructures, we perform projected density of states (pDOS) calculations for both Gr/WSSe/Gr and Gr/MnSSe/Gr heterostructures, projecting onto the states of carbon atoms, as well as onto individual atoms of the Janus monolayers. In both cases, see Figs.~\ref{fig:PROJWFC}(a) and \ref{fig:PROJWFC}(b), the pDOS confirms that the dominant states around the Fermi level belong to carbon atoms,
with negligible Janus monolayer atoms contribution near $E_{\rm F}$, suggesting the absence of hybridization between graphene and Janus monolayer states near the Fermi level.
Figures~\ref{fig:PROJWFC}(c) and \ref{fig:PROJWFC}(d) show 
the layer-resolved band structure of Gr/WSSe/Gr and 
Gr/MnSSe/Gr, respectively, in the vicinity of the graphene 
$K$ point. It reveals that the Dirac cones of the top and 
bottom graphene layers are fully decoupled and appear at 
different chemical potentials. This is a direct consequence of the intrinsic asymmetry of the Janus monolayers, in which the S and Se layers provide different environments for the top and bottom graphene layers, shifting their Dirac points to different energies.
We therefore conclude that the low-energy physics of both heterostructures can be described in terms of two independent graphene layers, each proximitized differently by its distinct surrounding and parametrized by a separate set of effective tight-binding parameters.

\section{Tight-binding model}\label{sec:TB}
The first-principles calculations presented above reveal that the low-energy 
electronic structure of both heterostructures can be described by two independent 
graphene layers. Since the crystallographic symmetry of both Gr/WSSe/Gr and 
Gr/MnSSe/Gr in the nonmagnetic phase is ${\bf C}_{3{\rm v}}$, we revisit the 
effective tight-binding model of graphene with ${\bf C}_{3{\rm v}}$ 
symmetry~\cite{Kochan2017},
\begin{equation}\label{Hgr}
    \mathcal{H}_{\rm Gr}({\bf C}_{3\rm v})=\mathcal{H}_\mathrm{orb}+\mathcal{H}_{\rm I}+\mathcal{H}_{{\rm R}},
\end{equation}
where $\mathcal{H}_\mathrm{orb}$ is the orbital Hamiltonian, $\mathcal{H}_{\rm I}$ 
is the intrinsic SOC term, and $\mathcal{H}_{{\rm R}}$ is the Rashba SOC term.

To this end, we define the effective $p_z$-orbital with spin $\sigma=\uparrow,\downarrow$ on sublattice $X=A,B$ and lattice site $m$ as $\ket{X_m\sigma}$. The first 
term in Eq.~\eqref{Hgr} represents the orbital Hamiltonian
\begin{eqnarray}
\mathcal{H}_\mathrm{orb}&=&\sum_{m,\sigma}\mu_A\ket{A_m\sigma}\bra{A_m \sigma}+\sum_{m,\sigma}\mu_B\ket{B_m\sigma}\bra{B_m \sigma}\nonumber\\
&&-t\sum_{<{m,n}>_{\rm nn}}\sum_{\sigma}\ket{X_m\sigma}\bra{X_n \sigma},
\end{eqnarray}
where $\mu_{A/B}=\mu\pm \Delta$ are the sublattice-dependent on-site potentials on sublattice $A$/$B$, with $\mu$ being the chemical potential and $\Delta$ the staggered potential. The second term describes the nearest-neighbor hopping, 
parametrized by the hopping strength $t$.

The intrinsic SOC term in Eq.~\eqref{Hgr} reads
\begin{equation}
    \mathcal{H}_{\rm I}=\sum_{\gamma={\rm A,B}}
    \sum_{\langle m,n\rangle_{\rm nnn},\sigma}
    \frac{{\rm i}\lambda_{\rm I}^{\gamma}}{3\sqrt{3}}
    \nu_{m,n}[s_z]_{\sigma\sigma}\ket{X_m\sigma}\bra{X_n \sigma},
\end{equation}
where the sum runs over next-nearest-neighbor pairs, 
$\lambda_{\rm I}^{{\rm A}/{\rm B}}$ are the sublattice-dependent 
intrinsic SOC parameters, and $\nu_{m,n}=\pm 1$ is a sign factor 
that takes the value $+1$ ($-1$) when the next-nearest-neighbor 
hopping from site $m$ to site $n$ via the common nearest neighbor 
encloses a clockwise (counterclockwise) path~\cite{Kochan2017}.

The last term in Eq.\eqref{Hgr} describes the nearest-neighbor Rashba SOC interaction  
\begin{equation}
\mathcal{H}_{\rm{R}}=\frac{2{\rm i}
\lambda_{\rm R}}{3}\sum_{<m,n>_{\rm nn}}^{\sigma\neq\sigma'}
[{\bf s}\times {\bf d}_{m,n}]_{\sigma\sigma'}^z \ket{X_m\sigma}\bra{X_n \sigma'},
\end{equation}
in which ${\bf s}$ is the vector of Pauli matrices, $\lambda_{\rm R}$ represents the Rashba SOC strength, while ${\bf d}_{m,n}$ is the unit vector in the horizontal plane pointing from lattice site $n$ to the nearest-neighbor site $m$. 

The Hamiltonian in Eq.~\eqref{Hgr} fully describes the low-energy 
physics of the proximitized top and bottom graphene layers in the 
Gr/WSSe/Gr heterostructure. In the case of Gr/MnSSe/Gr, the 
type-A antiferromagnetic ordering of MnSSe additionally induces 
opposite exchange interactions in the top and bottom graphene layers.  To account for this, we introduce the exchange-proximity Hamiltonian
\begin{equation}\label{Hex}
    \mathcal{H}_{\rm ex}=\Delta_{\rm A}\frac{\sigma_0+\sigma_z}{2}
    \otimes s_z+\Delta_{\rm B}\frac{\sigma_0-\sigma_z}{2}\otimes s_z,
\end{equation}
where $\Delta_{\rm A}$ and $\Delta_{\rm B}$ are the proximity-induced 
exchange parameters on sublattices $A$ and $B$, respectively, 
$s_z$ is the $z$-component of the spin Pauli matrix, and $\sigma_0$, 
$\sigma_z$ are the identity and $z$-component of the pseudospin 
Pauli matrices, respectively.

Finally, the effective Hamiltonians describing the top and bottom 
graphene layers in the two heterostructures are given by
\begin{eqnarray}
    \mathcal{H}_{\rm t/b}^{\rm Gr/WSSe/Gr}&=&\mathcal{H}_\mathrm{orb}
    +\mathcal{H}_{\rm I}+\mathcal{H}_{{\rm R}},\label{effW}\\ 
    \mathcal{H}_{\rm t/b}^{\rm Gr/MnSSe/Gr}&=&\mathcal{H}_\mathrm{orb}
    +\mathcal{H}_{\rm I}+\mathcal{H}_{{\rm R}}+\mathcal{H}_{\rm ex},\label{effMn}
\end{eqnarray}
where the subscript ${\rm t/b}$ refers to the top and bottom graphene 
layers, each parametrized by an independent set of tight-binding 
parameters extracted from first-principles calculations.
\section{Rashba Edelstein effect in bilayer graphene}\label{sec:charge-spin}
The proximity-induced SOC in graphene manifests itself experimentally through charge-to-spin conversion, where an applied charge current generates a transverse spin accumulation via the Rashba-Edelstein 
effect~\cite{E90,OMR+17}. The efficiency of this conversion is captured by the 
Rashba-Edelstein coefficient $\alpha_{\rm REE}$, which relates the 
induced spin accumulation to the applied charge current density. 
The sign of $\alpha_{\rm REE}$ is directly linked to the 
chirality of the Rashba-induced spin texture, since opposite chiralities give opposite signs of $\alpha_{\rm REE}$. This makes the Rashba-Edelstein coefficient an ideal quantity to detect the 
layer-selective chirality switch predicted in this work, as the top and bottom graphene layers are expected to exhibit opposite signs of $\alpha_{\rm REE}$, which can be independently accessed by gating.

For an electric bias applied to the system, represented by a field $E$, the non-equilibrium spin and current accumulation, $\delta \mathcal{O}$, $\mathcal{O}\in\{S_{y},J_{x}\}$, can be calculated using the linear response theory as
\begin{equation}\label{LRT}
\delta\mathcal{O}=\frac{E}{4\pi^2}\int d^2{\bm k}[\chi_{\mathcal{O}}^{\rm surf}({\bm k})+
     \chi_{\mathcal{O}}^{\rm sea}({\bm k})],
\end{equation}
where $S_{y}=\frac{\hbar}{2}s_{y}$, and $J_{x}=-ev_{x}$.
The $k$-dependent Fermi sea and Fermi surface response functions are expressed using the Kubo formula~\cite{K56,K57} in the Smr\v{c}ka-St\v{r}eda formulation~\cite{SS77,CB01,BM20}
\begin{eqnarray}
\chi_{\mathcal{O}}^{\rm surf}({\bm k})&=&
    \sum_{n,m} \frac{\hbar\gamma^2{\rm Re}\{
    \langle n{\bm k}|\mathcal{O}|m{\bm k}\rangle 
    \langle m{\bm k}|j_x|n{\bm k}\rangle \}}
    {\pi[(\epsilon_{n{\bm k}}-E_{\rm{F}})^2+\gamma^2][(\epsilon_{m{\bm k}}-E_{\rm{F}})^2+\gamma^2]}\nonumber\\
      \chi_{\mathcal{O}}^{\rm sea}({\bm k})&=&
    \sum_{n\neq m} 
    \frac{\hbar f_{n,m,{\bm k}}{\rm Im}\{
    \langle n{\bm k}|\mathcal{O}|m{\bm k}\rangle 
    \langle m{\bm k}|j_x|n{\bm k}\rangle\}}
    {(\epsilon_{n{\bm k}}-\epsilon_{m{\bm k}})^2} \label{eq:sea}
\end{eqnarray}
where $\epsilon_{n\bm{k}}$ are eigenenergies of the eigenstate $\ket{n\bm{k}}$ corresponding to the graphene Hamiltonian, and  $f_{n,m,\bf k{}}=f_{n,{\bf k}}-f_{m,{\bf k}}$ is the difference between Fermi-Dirac functions of eigenstates with eigenenergies $\epsilon_{n\bm{k}}$ and $\epsilon_{m\bm{k}}$  for the assumed 
electronic temperature of $k_{\rm B}T=0.01$~meV. 

The REE efficiency $\alpha_{\rm REE}$ is defined as the ratio of 
the non-equilibrium spin accumulation to the charge current response,
\begin{equation}\label{REEdef}
    \alpha_{\rm REE}=\frac{e v_{\rm{F}}}{\hbar}\frac{\delta S_y}{\delta J_x},
\end{equation}
where $\delta S_y$ is the current-induced nonequilibrium spin density along the $y$ axis, and both $\delta S_{y}$ and $\delta J_x$ are computed on equal footing from the same linear response formalism~\cite{K56,K57,SS77,CB01,BM20}.

Disorder scattering is treated phenomenologically through the 
broadening parameter~\cite{LSK+22,BM20,FBM14,ZZF+17} 
$\gamma=\hbar/2\tau$, where $\tau=10^{-10}$~s is the relaxation 
time~\cite{JLS+21}. The Fermi surface and Fermi sea contributions 
are integrated on a square grid around the $K$ and $K'$ points 
of size $0.006$~\AA$^{-1}$, with a discretization step of 
$\Delta k = 5\times10^{-7}$~\AA$^{-1}$.

\section{Results}\label{sec:results}

After establishing that the low-energy physics of both heterostructures 
can be described by two fully decoupled graphene layers, and 
defining the effective tight-binding models for Gr/WSSe/Gr and 
Gr/MnSSe/Gr in Eqs.~\eqref{effW} and~\eqref{effMn}, respectively, we proceed to extract the tight-binding parameters by fitting to the 
DFT band structure and spin expectation values.

\begin{figure}[t]
    \centering
\includegraphics[width=0.49\textwidth]{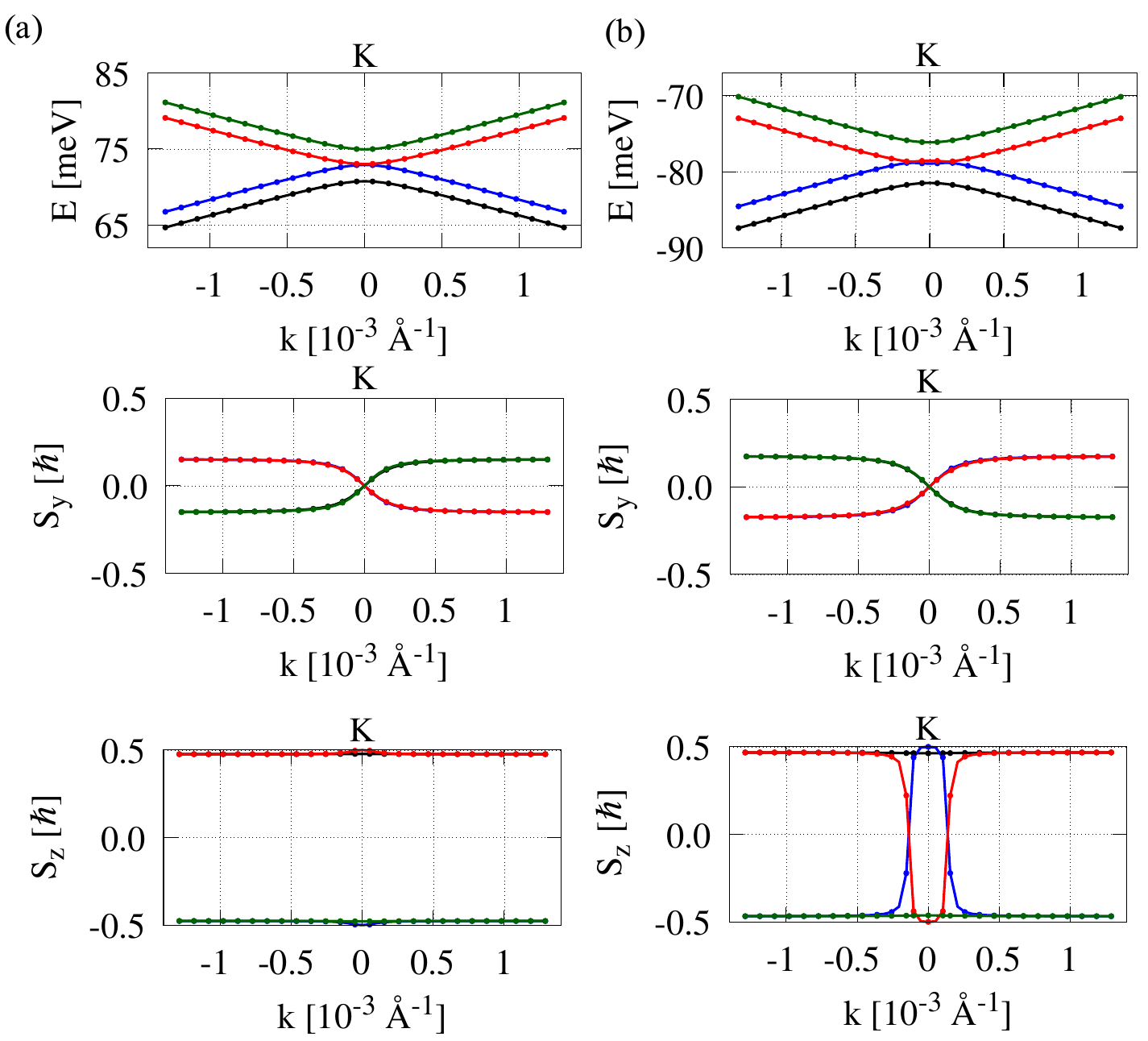}
    \caption{Comparison of the DFT electronic band structure and spin expectation values $S_y$ and $S_z$ near the Dirac $K$ point with the tight-binding model Hamiltonian~\eqref{effW} for the top (a) and bottom (b) graphene layers within the Gr/WSSe/Gr heterostructure. The tight-binding parameters are given in Table~\ref{TAB:parameters}. Solid lines represent the tight-binding fit, while circles denote DFT data. The $k$-path is taken along the $k_x$ direction ($k_y = 0$), centered at the $K$ point. By symmetry, $S_x$ vanishes along this path.}    \label{fig:WSSe}
\end{figure}

\begin{figure}[t]
    \centering
\includegraphics[width=0.499\textwidth]{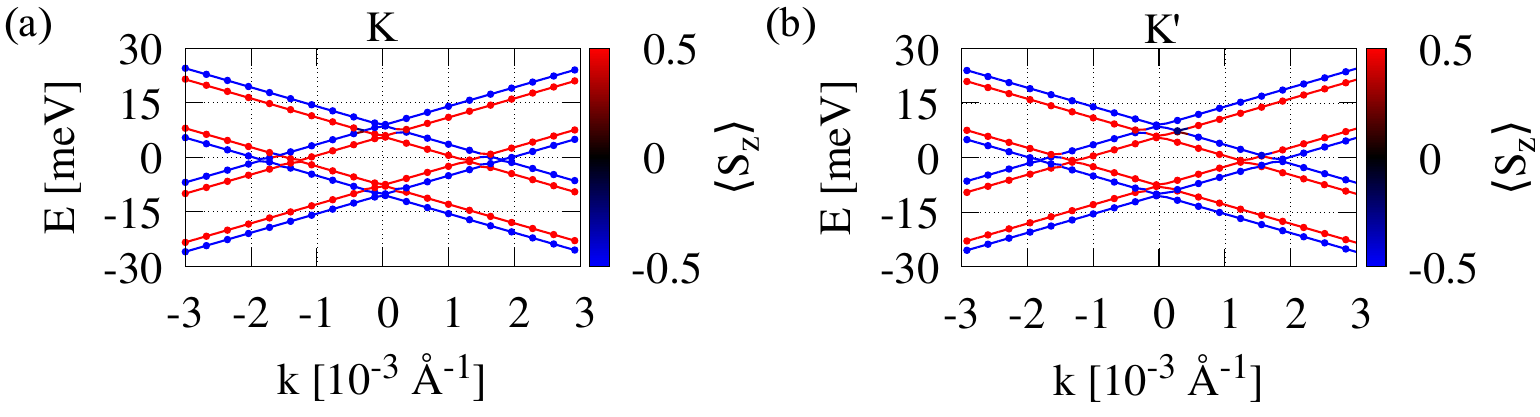}
    \caption{Comparison of the DFT electronic band structure and  $S_z$ spin expectation value near the Dirac $K$ (a) and $K'$ (b) points, with the tight-binding model Hamiltonian~\eqref{effMn} for the top and bottom graphene layers within the Gr/MnSSe/Gr heterostructure. The tight-binding parameters are given in Table~\ref{TAB:parameters}. Solid lines represent the tight-binding fit, circles denote DFT data, whereas the color scale compare the $S_z$ spin expectation values of the model with the DFT data. The $k$-path is chosen along the $k_x$ direction ($k_y = 0$), centered around $K$ and $K'$ point. The nearly identical graphs in (a) and (b) reflect the dominant role of the exchange interaction.
    We note that $S_x$ vanishes along this path. Nonzero $S_y$ (not shown here) spin expectation values suggest the presence of Rashba SOC, whose opposite signs in top and bottom graphene (see Table~\ref{TAB:parameters}) suggest opposite chirality of the in-plane spin texture.}    \label{fig:MnSSe}
\end{figure}

In the Gr/WSSe/Gr heterostructure case, the parameters are obtained by fitting the band 
structure and spin expectation values of both graphene layers in the 
vicinity of the $K$ point along the $k_x$ direction ($k_y=0$), 
within a fitting range of $0.0015$~\AA$^{-1}$ from the $K$ point.
In the Gr/MnSSe/Gr case, the fitting is performed simultaneously around 
both the $K$ and $K' (-K)$ points along the $k_y$ 
direction, within a fitting range of $0.003$~\AA$^{-1}$ from 
the $K$/$K'$ point. Fitting around both valleys simultaneously is essential to distinguish between the time-reversal symmetry-breaking 
proximity exchange interaction and time-reversal 
symmetry-conserving spin-orbit coupling, since these two 
contributions affect the $K$ and $K'$ points differently.
The extracted parameters are given in Table~\ref{TAB:parameters}, 
and the comparison between the DFT band structure and spin 
expectation values with the tight-binding model is shown in 
Figs.~\ref{fig:WSSe} and~\ref{fig:MnSSe} for Gr/WSSe/Gr and 
Gr/MnSSe/Gr, respectively.

\begin{table}[t]
\caption{Parameters of the effective tight-binding model of top and bottom graphene within the nonmagnetic, Gr/WSSe/Gr, and magnetic, Gr/MnSSe/Gr, heterostructure.}\label{TAB:parameters}
\centering
\footnotesize
\setlength{\tabcolsep}{7pt}
\renewcommand{\arraystretch}{1.0}
\begin{tabular}{cccccc}
\hline\hline
 Gr/WSSe/Gr  & $\mathcal{H}_{\rm t}$& $\mathcal{H}_{\rm b}$ \\ \hline
$t$\;[{\rm meV}]                       &2616.089 &2611.505\\
$\mu$\;[{\rm meV}]                     &72.897 &-78.763\\
$\Delta$\;[{\rm meV}]                  & 1.029 & 1.143\\
$\lambda_{\rm I}^{\rm A}$\;[{\rm meV}] & 0.924 & 1.302\\
$\lambda_{\rm I}^{\rm B}$\;[{\rm meV}] &-1.028 &-1.370\\
$\lambda_{\rm R}$\;[{\rm meV}]         &-0.309 &0.490\\\hline\hline
 Gr/MnSSe/Gr  & $\mathcal{H}_{\rm t}$& $\mathcal{H}_{\rm b}$ \\ \hline
$t$\;[{\rm meV}]                       &2398.291 &2400.830\\
$\mu$\;[{\rm meV}]&-9.017 &7.225\\
$\Delta$\;[{\rm meV}] &0.082 &0.016\\
$\lambda_{\rm I}^{\rm A}$\;[{\rm meV}] &-0.026 &0.002\\
$\lambda_{\rm I}^{\rm B}$\;[{\rm meV}] &-0.004 &-0.004\\
$\lambda_{\rm R}$\;[{\rm meV}] &-0.018&0.105\\
$\Delta_{\rm A}$\;[{\rm meV}] &1.281 &-1.511\\
$\Delta_{\rm B}$\;[{\rm meV}] &1.289 &-1.531\\\hline\hline
\end{tabular}
\end{table} 

We begin our analysis with the Gr/WSSe/Gr heterostructure. The first noticeable difference between the top and bottom graphene 
layers is a sizable offset between their chemical potentials, indicating a built-in electric field within the heterostructure that shifts the Dirac cones of the two layers in opposite energy directions. For spintronics applications, the most important effect is the opposite sign of the 
proximity-induced Rashba field in the top and bottom graphene layers. This sign reversal is a direct consequence of the inverted proximity 
geometry experienced by each graphene layer. Whereas the top graphene 
layer is proximitized from below by the WSSe monolayer, the bottom 
graphene layer is proximitized from above. Thus, each layer experiences 
an effective electric field pointing in the opposite direction. The 
difference in the absolute magnitude of the Rashba SOC between the 
two layers is a consequence of the inequivalent chemical environments 
imposed by the intrinsic structural asymmetry of the Janus WSSe 
monolayer.

We now turn to the Gr/MnSSe/Gr heterostructure. As in the 
Gr/WSSe/Gr case, we observe opposite chemical potential shifts 
and opposite signs of the Rashba SOC in the top and bottom 
graphene layers. However, as evident from both 
Fig.~\ref{fig:MnSSe} and Table~\ref{TAB:parameters}, the 
dominant energy scale of the spin-dependent parameters is set 
by the proximity exchange parameters $\Delta_{\rm A}$ and 
$\Delta_{\rm B}$, as expected given the magnetic nature of MnSSe. 
Specifically, each graphene layer acquires a ferromagnetic 
proximity exchange interaction of opposite sign, a direct 
consequence of the short-range nature of the proximity exchange: 
each graphene layer couples predominantly to its nearest Mn 
plane, and since the top and bottom Mn planes of MnSSe carry 
opposite magnetic orientations (see Fig.~\ref{fig:structure}(b)), 
opposite exchange fields are induced in the two graphene layers. 
Finally, we note that the SOC parameters are significantly 
smaller in Gr/MnSSe/Gr than in Gr/WSSe/Gr, reflecting the 
lighter atomic mass of Mn compared to W, which induces a much 
weaker proximity SOC in graphene.

\begin{figure}[t]
    \centering
\includegraphics[width=0.46\textwidth]{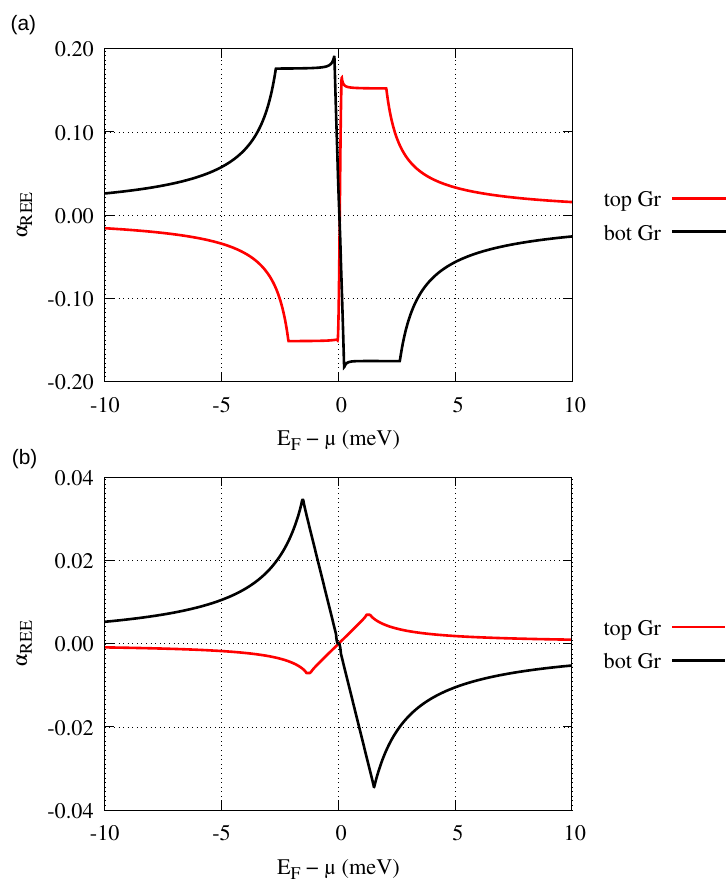}
    \caption{Calculated dependence of the conventional Rashba–Edelstein charge-to-spin conversion efficiency parameter $\alpha_{\rm REE}$ as a
function of doping for both top and bottom graphene within the (a) Gr/WSSe/Gr and (b) Gr/MnSSe/Gr heterostructures. }    \label{REE}
\end{figure}

The opposite signs of the proximity-induced Rashba SOC in the top 
and bottom graphene layers, see Table~\ref{TAB:parameters}, directly imply opposite chiralities of the in-plane spin texture, which should manifest experimentally as opposite signs of $\alpha_{\rm REE}$. To verify this, we calculate $\alpha_{\rm REE}$ as a function of doping, measured with respect to the chemical potential of each individual layer. The results are shown in Fig.~\ref{REE} for the top and bottom graphene layers of (a) Gr/WSSe/Gr and (b) Gr/MnSSe/Gr heterostructures. The $\alpha_{\rm REE}$ coefficients of the top and bottom graphene layers have opposite signs in both heterostructures, confirming the layer-selective chirality switch predicted from the tight-binding analysis. Furthermore, $|\alpha_{\rm REE}|$ of the bottom graphene layer exceeds that of the top graphene layer in both heterostructures, reflecting the stronger proximity-induced Rashba coupling experienced by the bottom layer, as quantified by the larger absolute value of $\lambda_{\rm R}$ in Table~\ref{TAB:parameters}.

The opposite sign of $\alpha_{\rm REE}$ in the top and bottom graphene layers has a direct physical consequence: the spin 
currents flowing in the top and bottom graphene layers point in opposite directions. This layer-selective chirality can be exploited as a spin current switch by independently activating source and drain electrodes on the top or bottom graphene layer via electrostatic gating. The studied heterostructures thus represent a promising platform for layer-selective chirality switching, conceptually analogous to, but mechanistically distinct from, the ferroelectric spin current switch recently proposed in graphene proximitized by In$_2$Se$_3$~\cite{MMJ+26}, where the 
chirality switch is achieved by reversing the ferroelectric polarization direction. In contrast, in the present heterostructures the chirality switch is inherent to the heterostructure geometry itself and is activated simply by 
selecting which graphene layer carries the current, offering a simpler and more direct route toward gate-tunable spintronic 
devices based on van der Waals heterostructures.

\section{Conclusions}\label{sec:conclusions}

We have investigated the electronic structure, spin textures, and charge-to-spin conversion in bilayer graphene intercalated by Janus 
nonmagnetic (WSSe) and magnetic (MnSSe) monolayers, using first-principles calculations, tight-binding modeling, and the  Kubo formalism.
We have shown that in both Gr/WSSe/Gr and Gr/MnSSe/Gr heterostructures, the Janus monolayer acts as a spacer 
that at the same time decouples the top and bottom graphene, while inducing proximity effects in each layer. The top-bottom asymmetry of the Janus structure breaks the horizontal 
mirror symmetry of AA-stacked bilayer graphene, shifts the Dirac cones of the two graphene layers to different chemical potentials, 
and, most importantly for spintronics, induces opposite signs of the proximity Rashba SOC in the top and bottom graphene layers. 
The opposite proximity Rashba fields in the top and bottom graphene 
layers lead directly to opposite signs of the Rashba-Edelstein 
efficiency $\alpha_{\rm REE}$ in the two layers. As a consequence, 
charge currents flowing in the top and bottom graphene layers 
generate spin accumulations of opposite sign, a layer-selective 
chirality switch that can be activated by independently gating the 
top or bottom graphene layer. We also note that while Gr/WSSe/Gr represents a realistic and 
experimentally accessible platform~\cite{Lin2020, Harris2023}, 
the Gr/MnSSe/Gr heterostructure serves as a model case 
demonstrating the generality of the layer-selective chirality 
switch mechanism in the presence of both proximity-induced SOC 
and out-of-plane exchange interaction. The latter is motivated 
by the experimental observation of out-of-plane antiferromagnetic 
ordering in the related monolayer MnSe~\cite{AHK+21}, and is 
further supported by our total energy calculations, which find 
an energy difference of only $0.19$~meV per Mn atom between the 
out-of-plane and in-plane spin configurations, confirming that 
the two are nearly degenerate. Moreover, proximity effects from 
the substrate materials have been shown to reorient the magnetic 
anisotropy in two-dimensional magnets~\cite{Pozsar2026}, suggesting that 
placing the Gr/MnSSe/Gr heterostructure on a suitable substrate 
could provide the additional magnetic anisotropy needed to 
stabilize the out-of-plane magnetic configuration experimentally.
Thus, bilayer graphene intercalated by Janus monolayers represents a promising platform for layer-selective spintronic devices, paving the way toward 
gate-tunable spin logic based on van der Waals heterostructures.

\section*{Acknowledgments}
\acknowledgments
Research results were obtained using the computational resources procured in the national project National competence centre for high performance computing (project code: 311070AKF2) funded by European Regional Development Fund, EU Structural Funds Informatization of society, Operational Program Integrated Infrastructure.
M.M. acknowledges the financial support by the EU NextGenerationEU through the Recovery and Resilience Plan for Slovakia under the Project No. 09I02-03-V01-00012, by the APVV grant APVV-23-0430, and VEGA grants 2/0081/26 and 2/0133/25.
\section*{Data Availability}
The data that support the findings of this study will be made 
publicly available on Zenodo upon acceptance of the manuscript.
\section*{Author Contributions}
M.M. conceived the study, performed all first-principles and 
tight-binding calculations, developed the charge-to-spin 
conversion code, analyzed all results, and wrote the manuscript.
\section*{Code Availability}
The charge-to-spin conversion calculations were performed using 
a code implementing the Kubo formalism defined in Eqs.~\eqref{LRT}-\eqref{REEdef} of this manuscript. 
The code is not publicly available.
\section*{Competing Interests}
The author declares no financial or non-financial competing 
interests.

\appendix
\section{Structural and computational details}\label{AppA}

\begin{table}[t]
\caption{The structural information of the studied Gr/WSSe/Gr and Gr/MnSSe/Gr heterostructures. The number $n$ describes the lattice vectors of the heterostructure $(n,0)=n{\bm a}_1$ and $(0,n)=n{\bm a}_2$ given in terms of the lattice vectors of the graphene lattice. The
applied strain $\chi$ to the graphene lattice parameter is also given, the total number of atoms $N$ in the heterostructure, and the number of carbon atoms on the top and bottom layer of graphene. Finally, unstrained lattice constants of WSSe and MnSSe are given.}\label{TAB:structural}
\centering
\footnotesize
\setlength{\tabcolsep}{7pt}
\renewcommand{\arraystretch}{1.0}
\begin{tabular}{cccccc}
\hline\hline
het  & $n$ &  $\chi$ [\%]& $N$ & $N_{\rm gr}^{\rm t/b}$ & $a$[\AA$]$\\\hline
Gr/WSSe/Gr & $4$ & -0.915 & 91 & 36/36 & 3.25~\cite{FGY+22,YSZ+23} \\\hline
Gr/MnSSe/Gr & $5$ & 2.195 & 137 & 50/50 & 4.19~\cite{SIC22} \\\hline
\end{tabular}
\end{table}

The lattice parameter of graphene is taken as $a_0=2.46$~\AA, while for Janus MnSSe and WSSe monolayers, the lattice parameters are equal to $a_{\rm MnSSe}=4.19$~\AA~\cite{SIC22} and $a_{\rm WSSe}=3.25$~\AA~\cite{FGY+22,YSZ+23}. The supercell is constructed by straining graphene, described by the relative strain $\chi$, using which the strained lattice parameter of graphene $a_0^{\rm str}$ can be described as $a_0^{\rm str}=(1+\chi[\%]/100)a_0$.  Using the strained lattice vectors of graphene ${\bm a}_1=a_0^{\rm str}{\bm e}_x$ and ${\bm a}_2=a_0^{\rm str}(\cos{(2\pi/3)}{\bm e}_x+\sin{(2\pi/3)}{\bm e}_y)$ we can define lattice vectors of the heterostructure $(n,0)$ and $(0,n)$ as $(n,0)=n{\bm a}_1$ and $(0,n)=n{\bm a}_2$. More detailed information of each heterostructure is
given in Table~\ref{TAB:structural}.

We performed the electronic structure calculation of Gr/WSSe/Gr and Gr/MnSSe/Gr heterostructures using DFT as implemented in the plane wave Q{\sc{uantum}} ESPRESSO~\cite{QE1,QE2} code. 

The relaxation of the Gr/WSSe/Gr heterostructure was performed using the Perdew-Burke-Ernzerhof functional~\cite{Perdew1996} and scalar-relativistic SG15 optimized norm-conserving Vanderbilt (ONCV) pseudopotentials~\cite{H13,SG15,SGH+16}. The kinetic energy cut-offs for the wave function and charge density were chosen to be 70~Ry and 280~Ry, respectively. Additionally, Methfessel–Paxton energy level smearing~\cite{MP89} of 1~mRy was used, and $6\times 6$ $k$-points mesh for the irreducible part of the Brillouin zone sampling were used for self-consistent calculations. 
The van der Waals interaction was modeled using the semiempirical Grimme-D2 correction~\cite{G06,Barone2009}, and a vacuum of 20~\AA~ in the $z$-direction to detach the periodic images of the heterostructure was used. The positions of atoms were relaxed using the quasi-Newton scheme using scalar-relativistic pseudopotentials, keeping the force and energy convergence thresholds for ionic minimization to $1\times10^{-4}$~Ry/bohr and $10^{-7}$~Ry, respectively. For the self-consistent calculation, including the spin-orbit coupling (with and without the applied perpendicular electric field), we use fully-relativistic ONCV pseudopotentials. Also, we have kept the same $k$-mesh but increasing the energy convergence thresholds to $10^{-8}$~Ry. Finally, dipole correction~\cite{B99} was applied to properly determine the Dirac point energy offset due to dipole electric field effects between graphene's and WSSe.

The relaxation of the Gr/MnSSe/Gr heterostructure was performed with the Perdew-Burke-Ernzerhof functional~\cite{Perdew1996} within the projector augmented-wave method~\cite{pslib,PAW}, using the kinetic energy cut-offs for the wave function and charge density 55~Ry and 326~Ry, respectively. 
Also, the same smearing, $k$-point mesh, force and energy convergence thresholds,  van der Waals interaction, relaxation scheme, and a vacuum in the $z$-direction was employed as in the Gr/WSSe/Gr case. In addition, to accurately describe strongly correlated Mn d orbitals, we set the on-site Hubbard U parameter to 2.3 eV~\cite{SIC22}. Perdew-Zunger exchange-correlation functional~\cite{PZ81} was used for self-consistent calculation in the relativistic case, with the dipole correction included, the same $k$-mesh, and the energy convergence thresholds increased to $10^{-8}$~Ry.
\section*{References}
\bibliography{biblio}
\end{document}